\documentclass[mypaper,10pt,twoside]{CoAst}
\usepackage{epsf,graphicx,fancyhdr}
\usepackage{amssymb}
\renewcommand{\chaptermark}[1]{\markboth{#1}{}}

\newcommand{\kopf}{\small\itshape Comm. in Asteroseismology\\ Vol. 143, 2003}
\newcommand{\Authors}[1]{\begin{center}\normalsize\bf\sf #1 \end{center}}

\renewcommand{\author}[1]{\begin{center}\normalsize\bf\sf #1 \end{center}}
\newcommand{\Address}[1]{\begin{center}\small\sf #1 \end{center}}

\renewenvironment{abstract}{\section*{Abstract}\normalsize\sf}{}
\newcommand{\References}[1]{\begin{flushleft}{\large References\\}\vspace*{2mm}\small #1 \end{flushleft}}

\newcommand{\chapterDSSN}[2]{\chapter[\sf\normalsize #1\\ \footnotesize \hspace*{5mm}by #2 \sf\normalsize][]{#1\\}\rhead[\fancyplain{}{\sf\footnotesize \center{#1}}]{\fancyplain{}{\sffamily\thepage}}\lhead[\fancyplain{\kopf}{\sffamily\thepage}]{\fancyplain{\kopf}{\sf\footnotesize \center{#2}}}}

\renewcommand{\chaptername}{}
\newcommand{\acknowledgments}[1]{\vspace*{5mm}\noindent\begin{bf}Acknowledgments. \end{bf} #1}

\begin{document}
\sf

\chapterDSSN{The convective envelope in $\gamma$~Doradus stars: theoretical uncertainties }
{J. Montalb\'an, A. Miglio, S. Th\'eado}

\Authors{J. Montalb\'an, A. Miglio and S. Th\'eado}
\Address{Institut d'Astrophysique et de G\'eophysique de l'Universit\'e de Li\`ege,
All\'ee du 6 Ao\^ut, 17 B-4000 Li\`ege, Belgium}

\noindent
\begin{abstract}
The depth of the convective envelope plays a fundamental role in the
driving mechanism proposed by Guzik et al. (2000) to explain the high-order g modes 
of $\gamma$~Dor pulsators.
In this poster we study the sensitivity of the convective envelope depth to
the description of convective transport,  to relevant physical processes, 
such as  microscopic diffusion, and to other uncertainties in theoretical stellar models.
\end{abstract}

\section{Depth of the convective envelope}

The ``convection blocking'' of radiation can drive high-order g-modes only for
stellar models with a temperature, at the bottom of the convection envelope (CE),
 between $2\times10^5$~K and $4.8\times10^5$~K (Guzik et al. 2000).
Unfortunately,  convection modeling is one of the most serious shortcomings in 
theoretical stellar evolution.
The ``standard model'' of convection, the mixing length theory (MLT), is a
simple model that contains essentially one adjustable parameter, 
 $\alpha$, which relates the mixing length to the local pressure scale height.
 Convection efficiency   increases with   $\alpha$ as well as, for 
a given stellar mass and chemical composition, the depth of the CE. 
Usually  $\alpha$  is  tuned to produce the solar radius at the solar age,
but  2D and 3D  numerical simulations of convection 
suggest that its value should decrease with 
increasing  stellar mass  so that, for the $\gamma$~Dor HR domain, it
should be lower than the solar value.
Furthermore, as the stellar mass increases, the effect of $\alpha$ on the stellar radius 
decreases, so that  1.5~M$_{\odot}$ stellar models computed with $\alpha$ between 1.8 and 1.4
have the same $T_{\rm eff}$  (i.e, corresponding to the middle of the observational $\gamma$~Dor 
instability strip, for a metal mass fraction Z=0.02) while the
depth of their CE is quite different (see Fig.1, left panel).

An alternative to the MLT is the Full Spectrum of Turbulence treatment of convection
(FST, Canuto et al.~1996).
 MLT is more efficient than FST in low efficiency convection regions, while
FST is much more efficient than MLT for highly
efficient convection. As a consequence,  the depth of CE
for FST models  changes from shallow to deep in a very narrow domain of $T_{\rm eff}$
(see Fig.~1 left panel). The range of $T_{\rm eff}$
of models whose $T_{\rm cz}$ is between $2\times10^5$~K and
$4.8\times10^5$~K is reduced with respect to the MLT case and, therefore, the width of the
$\gamma$~Dor instability strip predicted by FST treatment is also smaller.

The depth of the CE for models in the observational $\gamma$~Dor instability strip 
 is also affected by: 1. the microscopic diffusion, that increases
by He settling  the H abundance and, therefore, the opacity in the outer layers 
(see. Fig.~1, right panel), and that, by effect of  radiative acceleration
and consequent Fe accumulation, can produce
an additional convective region at  $2\times 10^5$~K.
2. the chemical composition: low metallicity models in the instability strip  have
shallower convective envelopes than solar metallicity ones.

\begin{figure*}
\centerline{
{\scriptsize
\begin{tabular}{c @{} c}
\includegraphics[width=0.75\textwidth]{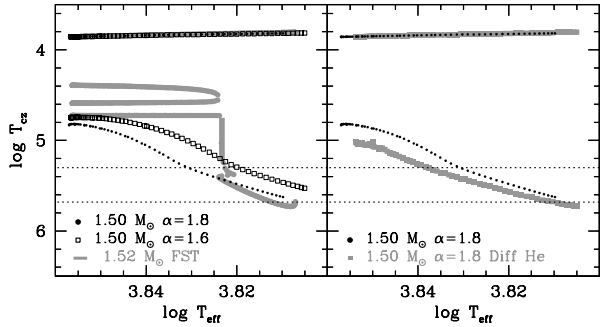} 
\end{tabular}}}
\par\vspace*{-4cm} 
\caption{\small Temperature  of the convective  envelope boundaries along the main sequence evolution of a
1.5~M$_{\odot}$ star. Left panel: for three different treatments of convection: MLT with $\alpha$=1.6 and 1.8, and
FST. Right panel: models without microscopic diffusion (dots) and with  gravitational settling of He (gray squares).}
\label{cores}  
\end{figure*}

\acknowledgments{
The authors acknowledge financial support from the Prodex-ESA Contract Prodex 8 COROT (C90199).
}

\References{
Guzik, J. A., Kaye, A. B., Bradley, P. A. et al.
 2000, ApJ 542, L57   \\

Canuto V.~M., Goldman I.,  Mazzitelli I., 1996, ApJ, 473, 550  \\

}

\end{document}